\documentclass{article}
\usepackage{ijcai25}

\usepackage{times}
\usepackage{soul}
\usepackage{url}
\usepackage[hidelinks]{hyperref}
\usepackage[utf8]{inputenc}
\usepackage[small]{caption}
\usepackage{graphicx}
\usepackage{amsmath}
\usepackage{amsthm}
\usepackage{booktabs}
\usepackage{algorithm}
\usepackage{algorithmic}
\usepackage{amssymb}
\usepackage{verbatim}
\usepackage{multirow}
\usepackage{multicol}
\usepackage{tikz}
\newcommand{\Circled}[1]{
	\begin{tikzpicture}[baseline=(char.base)]
		\node[inner sep=0pt,outer sep=0pt] (char) {\footnotesize{#1}};
		\draw circle (0.35em) node {};
	\end{tikzpicture}
}
\usepackage[switch]{lineno}

\title{Learning Neural Vocoder from Range-Null Space Decomposition}

\author{
	Andong Li$^{1, 2}$\and
	Tong Lei$^{3,4}$\and
	Zhihang Sun$^{3}$\and
	Rilin Chen$^{3}$\and
    Erwei Yin$^{5,6}$\and
	Xiaodong Li$^{1, 2}$\\
	Chengshi Zheng$^{1, 2}$\thanks{Chengshi Zheng is the corresponding author.}
	\affiliations
	$^1$Institute of Acoustics, Chinese Academy of Sciences\\
	$^2$University of Chinese Academy of Sciences \\ 
	$^3$Tencent AI Lab\\
	$^4$Nanjing University \\
	$^5$ Defense Innovation Institute, Academy of Military Sciences (AMS)\\
	$^6$ Tianjin Artificial Intelligence Innovation Center (TAIIC)
	\emails
	cszheng@mail.ioa.ac.cn
}

\begin{document}

\maketitle
\begin{abstract}
	Despite the rapid development of neural vocoders in recent years, they usually suffer from some intrinsic challenges like opaque modeling, and parameter-performance trade-off. In this study, we propose an innovative time-frequency (T-F) domain-based neural vocoder to resolve the above-mentioned challenges. To be specific, we bridge the connection between the classical signal range-null decomposition (RND) theory and vocoder task, and the reconstruction of target spectrogram can be decomposed into the superimposition between the range-space and null-space, where the former is enabled by a linear domain shift from the original mel-scale domain to the target linear-scale domain, and the latter is instantiated via a learnable network for further spectral detail generation. Accordingly, we propose a novel dual-path framework, where the spectrum is hierarchically encoded/decoded, and the cross- and narrow-band modules are elaborately devised for efficient sub-band  and sequential modeling. Comprehensive experiments are conducted on the LJSpeech and LibriTTS benchmarks. Quantitative and qualitative results show that while enjoying lightweight network parameters, the proposed approach yields state-of-the-art performance among existing advanced methods. Our code and the pretrained model weights are available at https://github.com/Andong-Li-speech/RNDVoC.
\end{abstract}

\section{Introduction}
Sound vocoder aims to reconstruct the audible time-domain waveforms using electronic and computational techniques, which is widely employed in text-to-speech (TTS)~{\cite{wang2017tacotron,huang2022fastdiff,li2025neural}}, text-to-audio (TTA)~{\cite{liu2023audioldm}}, and speech enhancement~{\cite{zhou2024mel,ijcai2022p582}}. Compared with traditional digital signal processing (DSP)-based vocoders like STRAIGHT~{\cite{kawahara2006straight} and WORLD~{\cite{morise2016world}}, neural vocoders enjoy superior advantage in generation naturalness and quality, thereby garnering wide attention in recent years.
\begin{figure}[t]
	\centering
	\centerline{\includegraphics[width=0.985\columnwidth]{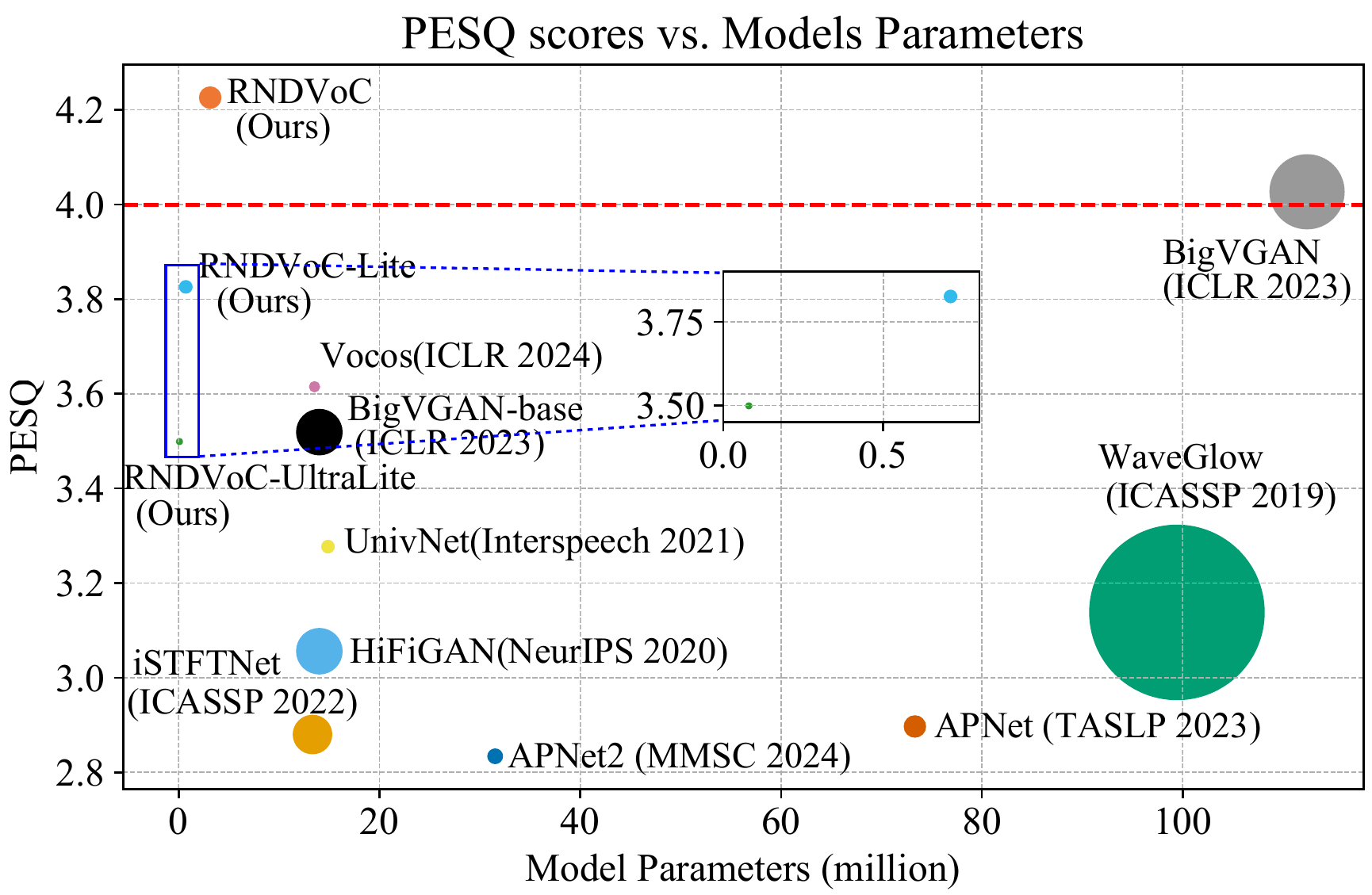}}
	\caption{A case comparison in terms of PESQ score on the LibriTTS benchmark. A larger bubble denotes higher computational complexity. Note that both BigVGAN and our methods are trained for 1M steps herein.}
	\label{fig:example}
\end{figure}

In the primitive stage, neural vocoders typically generate waveforms sample by sample, and representative works include WaveNet~{\cite{van2016wavenet}} and WaveRNN~{\cite{kalchbrenner2018efficient}}. Despite the improvement, they often suffer from considerably slow inference efficiency due to their autoregressive processing characteristic. To alleviate the deficiency, some other schemes have subsequently been adopted, including knowledge distillation-based~{\cite{oord2018parallel}}, normalization flow-based~{\cite{ping2020waveflow}}, and glottis-based~{\cite{juvela2019glotnet}}. Recently, generative adversarial network (GAN)-based neural vocoders have gained increasing attention due to their non-autogressive structure and high generation quality. Pioneering works include MelGAN~{\cite{kumar2019melgan}} and HiFiGAN~{\cite{kong2020hifi}}, where the mel-spectrograms are gradually upsampled and recovered to waveforms via alternate residual blocks and upsampling layers. Except for the time domain based methods, time-frequency (T-F) domain based neural vocoders have been proposed in more recent years, which enjoy faster inference speed and promising generation quality~{\cite{siuzdakvocos}}.
	
Despite the success of existing neural vocoders, some inherent challenges still exist and can impede the further progress. First, most of the end-to-end neural vocoders still suffer from the typical parameter-performance trade-off. For example, although achieving impressive performance, the number of parameters of BigVGAN is as high as 112M~{\cite{leebigvgan}}, which is awkward and also inconvenient for practical applications. Second, while T-F domain based neural vocoders exhibit the advantage in inference efficiency, their performance often lags behind mainstream time-domain-based ones. For the first point, we argue that existing methods usually model the vocoder process as a black-box, neglecting the utilization the linear degradation prior of mel-spectrogram{\footnote{Strictly speaking, mel-spectrogram is linearly compressed in the spectral magnitude, and the phase information is dropped.}}. Therefore, excessive parameters are often required for overall target reconstruction. For the second point, short-time Fourier transform (STFT) operation explicitly decouples the frequency information among different sub-bands, and acoustic features usually lie in specific frequency regions. However, only full-band modeling is adopted in existing T-F domain based neural vocoders, neglecting the hierarchical characteristic of the T-F spectrum.
	
To this remedy, we propose a novel T-F domain based neural vocoder called $\textbf{RNDVoC}$ to tackle the above-mentioned challenges. Specifically, we revisit the formulation of mel-spectrogram and establish the connection with range-null decomposition (RND) theory. By doing so, the reconstruction of target spectrum is thus decomposed into range-space modeling (RSM) and null-space modeling (NSM), where the former aims to project the original mel-spectrogram into the corresponding counterpart defined in the target linear-scale domain, and the latter is responsible for spectral details generation. In this way, we provide a more transparent and also elegant perspective for framework design. Besides, we devise an efficient dual-path structure, where the spectrum is hierarchically encoded and decoded, and the cross-band and narrow-band modules are alternate adopted for sub-band and sequential modeling. Experimental results show that with only only \textbf{2.8\%} parameters and \textbf{8.17\%} computational complexity, our method yields comparable and even better performance over BigVGAN-112M version in both objective and subjective scores, as well as nearly 10$\times$ speed-up on a CPU. Meanwhile, when the network parameters is further reduced as low as \textbf{0.08M}, the performance is still comparable to existing baselines. Figure~{\ref{fig:example}} showcases an example of the PESQ scores using many existing representative methods on the LibriTTS benchmark. Our contributions can be summarized as four-fold:
\begin{itemize}
	\item We provide a novel range-null space decompostion perspective to tackle the neural vocoder task. To our best knowledge, this is the first time to leverage the RND theory for audio generation.
	\item We devise a novel dual-path framework to encode/decode the spectrum hierarchically, and cross-band and narrow-band modules are employed for efficient sub-band and sequential modeling.
	\item We conduct extensive experiments to reveal the superiority of our method over existing baselines.
	\item We provide an ultra-light version with only around 80K trainable parameters. To our best knowledge, this is the smallest end-to-end neural vocoder up to now. 
\end{itemize} 
\begin{figure}[t]
	\centering
	\centerline{\includegraphics[width=0.98\columnwidth]{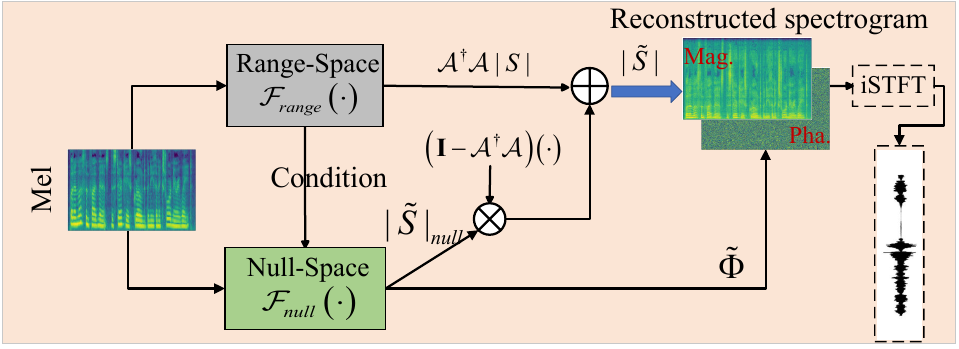}}
	\caption{Illustrations of the proposed RNDVoC.}
	\label{fig:system-diagram}
\end{figure}
\begin{figure*}[t]
	\centerline{\includegraphics[width=0.99\textwidth]{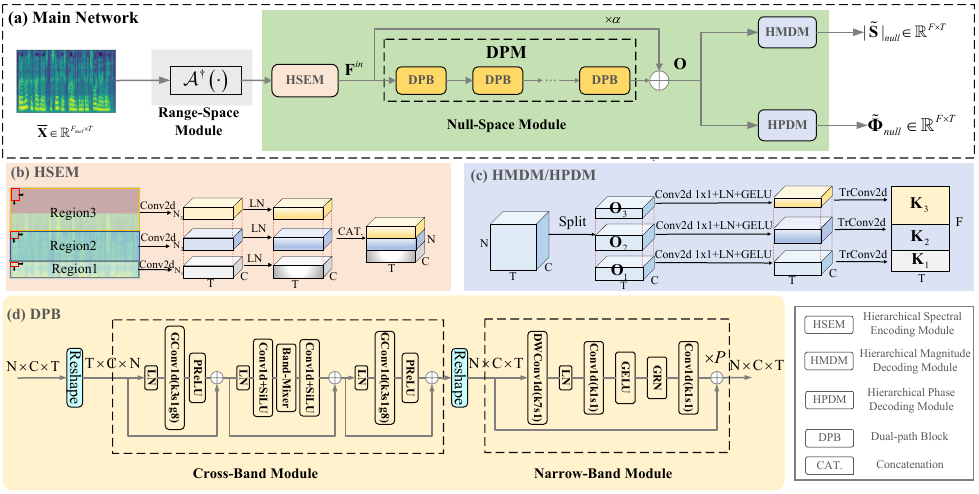}}
	\caption{Network structure of the proposed RNDVoC. (a) Overall network structure of the RNDVoC. (b) Detailed structure of the hierarchical spectral encoding module (HSEM). (c) Detailed structure of the hierarchical magnitude decoding module (HMDM) and hierarchical phase decoding module (HPDM). (d) Detailed structure of the dual-path block (DPB). Diffferent modules are indicated with different colors.}
	\label{fig:framework-detail}
\end{figure*}
\section{Related Works}\label{sec:related-work}
%In this section, we will briefly retrospect DSP- and neural-based vocoders. 
\subsection{DSP-based Methods}
In conventional DSP-based vocoders, the speech is usually generated with statistical parameter estimation. In STRAIGHT~{\cite{kawahara2006straight}}, the excitation and resonant parameters are separately estimated for real-time speech manipulation and synthesis. In~{\cite{morise2016world}}, the F0, spectral envelop, and aperiodic parameters are first determined using analysis algorithms. A synthesis method is subsequently devised for the time-domain waveform generation. Although being simplicity, the synthesized speech quality is often unsatisfactory and may include some buzzing artifacts.
\subsection{Neural Vocoder Methods}

\textbf{Autoregressive methods}: In WaveNet~{\cite{van2016wavenet}}, consecutive WaveNet modules with increasing dilation rates are adopted for sample-level generation. WaveRNN~{\cite{kalchbrenner2018efficient}}, samples are autoregressively generated with RNN layers. In LPCNet~{\cite{valin2019lpcnet}}, the linear prediction coefficients (LPC) are estimated to predict the next sample and a lightweight RNN is utilized for residual calculation. Despite the improvements over traditional DSP-based vocoders, the autoregressive nature can suffer from rather slow inference efficiency.\\
\textbf{Flow-based methods}: Normalization flow is regarded as a classical generation paradigm. Typical flow-based neural vocoders include WaveGlow~{\cite{prenger2019waveglow}}, FlowWaveNet~{\cite{kim2019flowavenet}}, and RealNVP~{\cite{dinh2016density}}, where the bijective mapping is established between a normalization probability distribution and target data distribution through stacked invertible modules.\\
\textbf{GAN-based methods}: In MelGAN~{\cite{kumar2019melgan}}, the mel-spectrogram is gradually upsampled through consecutive upsampling layers and residual blocks. In HiFiGAN, to facilitate the periodic pattern generation, the multi-periodic and multi-scale discriminators are employed for improved speech generation. BigVGAN~{\cite{leebigvgan}} incorporates the periodic activation function and anti-aliased representation into the generator to boost the periodic components generation, and the network size is further scaled up to 112M for universal audio vocoding. Except for the above-mentioned time-domain methods, several works explore the feasibility in the T-F domain. Vocos~{\cite{siuzdakvocos}} stacks multiple ConvNext v2 blocks~{\cite{woo2023convnext}} for magnitide and phase estimation. In~{\cite{ai2023apnet}}, the magnitude and phase components are separately modeled with ResNet blocks.\\
\textbf{Diffusion-based methods}: Owing to the powerful generation capability of duffusion methods, some works also adopt diffusion for neural vocoder. Representative works include DiffWave~{\cite{kongdiffwave}}, WaveGrad~{\cite{chenwavegrad}}, and PriorGrad~{\cite{leepriorgrad}}, where the waveforms gradually degrade to Gaussian noise in the forward path and the target signals can be iteratively generated in the reverse process. Despite good performance, the implemention efficiency can be slow due to numerous iteration steps, and fast sampling strategies~{\cite{huang2022prodiff,lu2022dpm}} are required to reduce the inference cost.

\section{Method}\label{sec:method}
In this section, we first introduce the range-null decomposition (RND) theory and formulate the problem of the speech vocoder task, and then we demonstrate their interconnection. Finally, the proposed learning framework is elucidated.
\subsection{Range-Null Space Decomposition}\label{sec:range-null-space}
For a classical signal compression physical model:
\begin{align}
	\label{eqn:1}
	\mathbf{y} = \mathbf{A}\mathbf{x} + \mathbf{n},
\end{align}
where $\mathbf{x}\in\mathbb{R}^{D}$, $\left\{\mathbf{y}, \mathbf{n}\right\}\in\mathbb{R}^{d}$ denote the target, observed, and noise signals, respectively, and $\mathbf{A}\in\mathbb{R}^{d\times D}$ denotes the compression matrix, with $d\ll D$. In the noise-free scenario, Eq.({\ref{eqn:1}}) can be simplified into $\mathbf{y} = \mathbf{A}\mathbf{x}$. If the pseudo-inverse of $\mathbf{A}$ is defined as $\mathbf{A}^{\dagger}\in\mathbb{R}^{D\times d}$, which satisfies $\mathbf{A}\mathbf{A}^{\dagger}\mathbf{A}\equiv\mathbf{A}$, then the signal $\mathbf{x}$ can be decomposed into two orthogonal sub-spaces: one residing in the range-space of $\mathbf{A}$, and the other in the null-space:
\begin{align}
	\label{eqn:2}
	\mathbf{x}\equiv \mathbf{A}^{\dagger}\mathbf{A}\mathbf{x} + \left(\mathbf{I} - \mathbf{A}^{\dagger}\mathbf{A}\right)\mathbf{x},
\end{align}
where $\mathbf{A}^{\dagger}\mathbf{A}\mathbf{x}$ defines the range-space component and $\left(\mathbf{I} - \mathbf{A}^{\dagger}\mathbf{A}\right)\mathbf{x}$ corresponds to the remaining null-space component. For practical solution of $\mathbf{x}$, \emph{i.e.}, $\tilde{\mathbf{x}}$, two consistency conditions should be satisfied: (1) Degradation consistency: $\mathbf{A}\mathbf{\tilde{x}}\equiv\mathbf{y}$, where the original information embedded in the compressed signal space remains unaltered after reconstruction. (2) Data consistency: $\mathbf{\tilde{x}}\sim p\left(\mathbf{x}\right)$, where the estimation $\tilde{\mathbf{x}}$ should share the same data distribution to $\mathbf{x}$. Based on the above constraints, the solution can be expressed as:
\begin{align}
	\label{eqn:3}
	\tilde{\mathbf{x}} = \mathbf{A}^{\dagger}\mathbf{y} + \left(\mathbf{I} - \mathbf{A}^{\dagger}\mathbf{A}\right)\hat{\mathbf{x}},
\end{align}
where $\mathbf{A}^{\dagger}\mathbf{y}$ and $\left(\mathbf{I} - \mathbf{A}^{\dagger}\mathbf{A}\right)\hat{\mathbf{x}}$ correspond to the solutions in the range-space and null-space, respectively. Due to the orthogonality characteristic, $\tilde{\mathbf{x}}$ naturally satisfies the first condition, \emph{i.e.}, $\mathbf{A}\tilde{\mathbf{x}}\equiv\mathbf{A}\mathbf{A}^{\dagger}\mathbf{A}\mathbf{x} + \mathbf{A}\left(\mathbf{I} - \mathbf{A}^{\dagger}\mathbf{A}\right)\hat{\mathbf{x}}\equiv\mathbf{A}\mathbf{x}+\mathbf{0}\equiv\mathbf{y}$. Therefore, we only need to consider the estimation of $\hat{\mathbf{x}}$ to meet the second condition, which can often be implemented with learnable networks. 
\subsection{Revisit Neural Vocoder Tasks}\label{sec:revisit-neural-vocoder-task}
In this paper, we mainly consider the scenario when the mel-spectrogram serves as the acoustic feature due to its simplicity and commonality in neural vocoders. The degradation process of a log-scale mel-spectrogram can be given by:
\begin{align}
	\label{eqn:4}
	\mathbf{X}^{mel} =\log\left(\mathcal{A}|\mathbf{S}|\right),
\end{align}
where $\mathbf{X}^{mel}\in\mathbb{R}^{F_{m}\times T}$ and $|\mathbf{S}|\in\mathbb{R}^{F\times T}$ denote the mel and target spectrograms, respectively, $\left\{F_{m}, F\right\}$ are the frequency size in the mel- and linear-scale, $T$ is the frame size. $\mathcal{A}\in\mathbb{R}^{F_{m}\times F}$ refers to the mel-filter, which is instantiated by a linear compression matrix. 

It is evident that Eq.~({\ref{eqn:4}) involves two degradation operations: $\Circled{\footnotesize{1}}$ Discarding the phase information, $\Circled{\footnotesize{2}}$ Magnitude compression with a linear operation. Intuitively, two steps are necessarily involved in the inverse process:  (1) Estimating the phase spectrum, (2) Recovering the magnitude spectrum from the compressed observation.
	
In the preliminary literature, a neural network typically serves as the black-box to establish the mapping relation between target waveform/T-F spectrum and mel feature, which can be expressed as:
\begin{align}
	\label{eqn:5}
	\tilde{\mathbf{s}} = \mathcal{F}_{T}\left(\mathbf{X}^{mel}, \Theta_{1}\right), \left\{|\tilde{\mathbf{S}}|, \tilde{\mathbf{\Phi}}\right\} = \mathcal{F}_{TF}\left(\mathbf{X}^{mel}, \Theta_{2}\right),
\end{align}
where $\mathcal{F}_{T}\left(\cdot,\Theta_{1}\right)$ and $\mathcal{F}_{TF}\left(\cdot,\Theta_{2}\right)$ refer to the mapping function of time-domain and T-F domain based neural vocoders, respectively. $\tilde{\mathbf{s}}$ is the estimated time-domain waveform, and $\left\{|\tilde{\mathbf{S}}|, \tilde{\mathbf{\Phi}}\right\}$ denote the estimated spectral magnitude and phase, respectively. Note that if the log-operation is absorbed into the left of Eq.~({\ref{eqn:4}), \emph{i.e.}, $\overline{\mathbf{X}}^{mel} = \exp\left(\mathbf{X}^{mel}\right)=\mathcal{A}|\mathbf{X}|$, it then exhibits a similar format to the noise-free case of Eq.~({\ref{eqn:1}). In other words, the original magnitude spectrum recovery can be formulated into a classical compressive sensing (CS) problem~{\cite{baraniuk2010model}}, in which we attempt to reconstruct the target signal from a linearly-compressed representation. Following the explicit decomposition in Eq.~({\ref{eqn:3}), the abstract generation process of the target spectrum in Eq.~({\ref{eqn:5}}) can be rewritten as:
\begin{align}
	\label{eqn:6}
	|\tilde{\mathbf{S}}|_{range} = \mathcal{F}_{range}\left(\mathbf{\overline{X}}^{mel}\right) = \mathcal{A}^{\dagger}\mathbf{\overline{X}}^{mel},
\end{align}
\begin{align}
	\label{eqn:7}
	\left\{|\tilde{\mathbf{S}}|_{null}, \tilde{\mathbf{\Phi}}\right\} = \mathcal{F}_{null}\left(|\tilde{\mathbf{S}}|_{range}\right),
\end{align}
\begin{align}
	\label{eqn:8}
	|\tilde{\mathbf{S}}| &= |\tilde{\mathbf{S}}|_{range} + \left(\mathbf{I} - \mathcal{A}^{\dagger}\mathcal{A}\right)|\tilde{\mathbf{S}}|_{null}\nonumber\\
	&= \mathcal{A}^{\dagger}\mathcal{A}\left|\mathbf{S}\right| + \left(\mathbf{I} - \mathcal{A}^{\dagger}\mathcal{A}\right)|\tilde{\mathbf{S}}|_{null},
\end{align}
\begin{align}
	\label{eqn:9}
	\tilde{\mathbf{S}} = |\tilde{\mathbf{S}}|e^{j\mathbf{\tilde{\Phi}}},
\end{align}
where $\mathcal{F}_{range}\left(\cdot\right)$ and $\mathcal{F}_{null}\left(\cdot\right)$ are the operations in the range- and null-spaces, respectively, $\mathcal{A}^{\dagger}\in\mathbb{R}^{F\times F_{m}}$ is the pseudo-inverse of $\mathcal{A}$. As shown in Figure~{\ref{fig:system-diagram}}, in $\mathcal{F}_{range}$, the original acoustic feature in the mel-domain is transformed into the target domain via the pseudo-inverse operation. In $\mathcal{F}_{null}$, the remaining magnitude details are estimated to supplement the overall spectral reconstruction. Compared with preliminary methods, the reconstruction format in Eq.~({\ref{eqn:8}}) enjoys several advantages: First, we fully utilize the linear degradation of mel-spectrum as the prior to establish two orthogonal sub-spaces, which enhances the overall framework interpretability. Besides, due to the degradation consistency, the original acoustic feature embedded in the mel-spectrogram can be well preserved, mitigating the acoustic distortion in the reconstructed signals. One should note that as only the spectral magnitude is involved in degradation formulation, we enforce $\mathcal{F}_{null}$ to also estimate the phase component, as shown in Eq.~({\ref{eqn:7}). Besides, we notice that most recently, a similar pseudo-inverse strategy is adopted in~{\cite{lv24_interspeech}}. The difference is that here we endow the pseudo-inverse operation with a more intuitive physical explanation while it is only adopted as an empirical trick in~{\cite{lv24_interspeech}}.
\subsection{Architecture Design of RNDVoC}\label{sec:architecture-design-of-rndvoc}
As previously illustrated, we explicitly decouple two orthogonal sub-spaces and devise the network modules for estimation. Figure~{\ref{fig:framework-detail}}(a) presents detailed framework of the RNDVoC. Given the input mel-spectrogram, it is processed by the range-space module (RSM), where the pseudo-inverse operation is adopted to shift the mel into the linear-scale domain. After that, the null-space module (NSM) is utilized, and three modules are mainly involved: hierarchical spectral encoding module (HSEM), dual-path module (DPM), and two decoder branches, \emph{i.e.}, hierarchical magnitude/phase decoding module (HMDM/HPDM), which will be introduced in a sequel.
\begin{figure}
	\centering
	\includegraphics[width=0.47\textwidth]{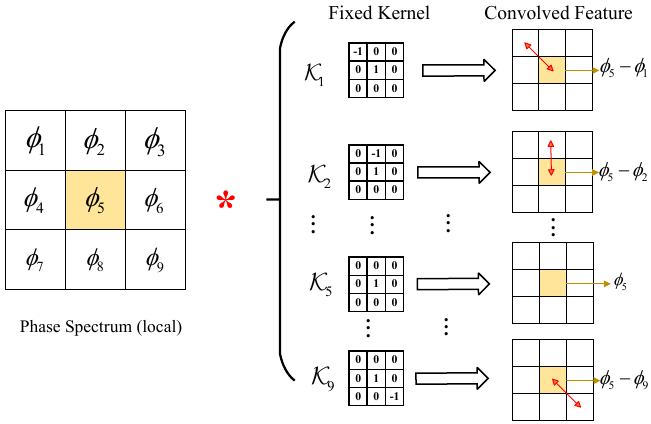}
	\caption{Illustration of the proposed omnidirectional phase loss.}  
	\label{fig:omni_loss}
\end{figure}
\renewcommand\arraystretch{1.20}
\begin{table*}
	\centering
	\Huge
	\resizebox{0.985\textwidth}{!}{
		\begin{tabular}{cc|cccc|ccccccc}
			\toprule
			\multirow{2}*{Models} &\multirow{2}*{Domain} &\#Param. &\#MACs &\multicolumn{2}{c|}{Inference Speed} &\multirow{2}*{M-STFT$\downarrow$} &\multirow{2}*{PESQ$\uparrow$} &\multirow{2}*{MCD$\downarrow$} &Periodicity$\downarrow$ &V/UV$\uparrow$ &Pitch$\downarrow$ &\multirow{2}*{VISQOL$\uparrow$}\\
			\cline{5-6}
			& &(M) &(Giga/5s) &CPU &GPU & & & &RMSE &F1 &RMSE & \\
			\midrule
			HiFiGAN-V1 &T &13.94 &152.90 &0.1669(5.99$\times$) &0.0069(145$\times$) &1.1699 &3.574 &3.6711 &0.1344 &0.9474 &33.6874 &4.7706 \\
			iSTFTNet-V1 &T &13.26 &107.76 &0.0949(10.54$\times$) &0.0038(266$\times$) &1.1883 &3.535 &3.6252 &0.1356 &0.9466 &35.1055 &4.7557 \\
			Avocodo &T &13.94 &152.95 &0.1611(6.21$\times$) &0.0063(158$\times$) &1.1653 &3.604 &3.6570 &0.1386 &0.9462 &32.9887 &4.7714\\
			BigVGAN-base &T &13.94 &152.90 &0.3856(2.59$\times$) &0.0368(27$\times$) &0.9784 &3.603 &2.3314 &0.1198 &0.9562 &30.2774 &4.8217 \\
			BigVGAN &T &112.18 &417.20 &0.6674(1.50$\times$) &0.0507(20$\times$) &\textbf{0.9001} &\textbf{4.107} &\textbf{1.8769} &\textbf{0.0838} &\textbf{0.9716} &\textbf{20.6922} &\textbf{4.8699}\\
			APNet &T-F &72.19 &31.11 &0.03(33.33$\times$) &0.0022(458$\times$) &1.2659 &3.390 &3.2847 &0.1508 &0.9454 &23.0571 &4.6947\\
			APNet2 &T-F &31.38 &13.53 &0.0187(53.48$\times$) &0.0016(643$\times$) &0.9815 &3.492 &2.8288 &0.1126 &0.9592 &25.3629 &4.7515 \\
			FreeV &T-F &18.19 &\underline{7.84} &\underline{0.0139(71.94$\times$)} &\underline{0.0011(951$\times$)} &1.0076 &3.593 &2.7502 &0.1118 &0.9603 &25.9922 &4.7427\\
			Vocos &T-F &13.46  &\textbf{5.80} &\textbf{0.0066(151.52$\times$)} &\textbf{0.0007(1400$\times$)} &1.0123 &3.522 &2.6699 &0.1213 &0.9559 &29.1304 &4.7741\\
			\hline
			\textbf{RNDVoC(Ours)} &T-F &\textbf{3.14} &34.10 &0.0669(14.96$\times$) &0.0043(233$\times$) &\underline{0.9103} &\underline{3.987} &\underline{2.0471} &\underline{0.0854} &\underline{0.9714} &\underline{21.3183} &\underline{4.8373}\\
			\hline
	\end{tabular}}
	\caption{Objective comparisons among different baselines on the LJSpeech benchmark. $\text{T}$ and $\text{T-F}$ refer to time-domain and T-F domain-based methods, respectively. ``$a\times$'' denotes the speed-up ratio over real-time. The inference speed on a CPU is evaluated based on a CPU Intel(R) Core(TM) i7-14700F. For GPU, it is based on NVIDIA GeForce RTX 4060 Ti. The best and second-best performances are respectively highlighted in \textbf{bold} and \underline{underlined}, respectively.}
	\label{tbl:objective-metric-ljs}
\end{table*}
\subsubsection{Hierarchical Spectral Encoder and Decoder}
Detailed network structure of the HSEM is shown in Figure~{\ref{fig:framework-detail}}(b). Given the input $|\tilde{\mathbf{S}}|_{range}\in\mathbb{R}^{F\times T}$, it is first split into $I$ regions, and the $i$-th spectral region is notated as $|\tilde{\mathbf{S}}|_{range,i}\in\mathbb{R}^{F_{i}\times T}$. As the major acoustic features lie in the low/mid frequency regions (\emph{e.g.} fundamental frequency F0), we follow the ``from-fine-to-coarse'' principle for region division, that is, we gradually increase $F_{i}$ with the increase in region index. For each spectral region, a separate Conv2d with stride is then utilized for spectral compression, followed by a layer-normalization (LN) layer~{\cite{lei2016layer}}. The process can be formulated as:
\begin{align}
	\label{eqn:10}
	\mathbf{F}^{in}_{i} = \text{LN}\left(\text{Conv2D}\left(|\tilde{\mathbf{S}}|_{range,i}\right)\right)\in\mathbb{R}^{N_{i}\times C\times T},
\end{align}
where $\mathbf{F}^{in}_{i}$ is the compressed spectral feature of the $i$-th region, and $\left\{N_{i}, C\right\}$ denote the sub-band and channel size, respectively. Here $C$ is set to 256. Similarly, for the decoding process, two branches are utilized for magnitude and phase estimation, respectively. Taking the magnitude branch as an example, detailed structure is shown in Figure~{\ref{fig:framework-detail}}(c). Given the input feature $\mathbf{O}\in\mathbb{R}^{N\times C\times T}$, it is first split into $I$ regions, \emph{i.e.}, $\left\{\mathbf{O}_{1},\cdots,\mathbf{O}_{I}\right\}$. For each region, we first adopt a separate point-wise Conv2d layer, followed by LN and GELU activation layer. After that, a TrConv2d is utilized to recover the spectral target. The formulation can be expressed as:
\begin{equation}
	\label{eqn:11}
	\mathbf{K}_{i} = \text{TrConv2d}\left(\text{GELU}\left(\text{LN}\left(\text{Conv2d1x1}\left(\mathbf{O}_{i}\right)\right)\right)\right).
\end{equation}

For the magnitude branch, the $\exp\left(\cdot\right)$ is applied to guarantee the non-negativity. For the phase branch, $\mathbf{K}_{i}$ is split into real and imaginary components, and $\text{Atan2}\left(\cdot\right)$ is adopted for phase calculation. Note that in~{\cite{yu23b_interspeech}}, a similar band-split strategy is adopted, where the spectrum is split into $N$ non-uniform sub-bands, and each sub-band is separately encoded and decoded. The difference is that here the number of regions $I$ is usually far less than $N${\footnote{In our practical settings, $I = 3$, and $N = 24$.}}, and thus fewer loop operations are required, leading to faster inference speed. Besides, as the network weights within one spectral region are shared, fewer trainable parameters are needed. 
\subsubsection{Dual-Path Block}\label{sec:dual-path-block}
To facilitate the spectral information modeling, $B=6$ dual-path blocks (DPBs) are stacked, each of which is shown in Figure~{\ref{fig:framework-detail}}(d). Specifically, it consists of a cross-band module and a narrow-band module, namely corresponding to the time and sub-band modeling. 

Given the input feature $\mathbf{F}^{\left(b\right)}\in\mathbb{R}^{N\times C\times T}$, it is first reshaped and sent to the cross-band module (CBM). Motivated by~{\cite{quan2024spatialnet}}, we adopt a light-weight design for cross-band modeling, and it involves three steps. First, a LN, followed by a group convolution with kernel being 3 along the sub-band axis and PReLU~{\cite{he2015delving}}, is adopted to model the correlation between neighboring sub-bands:
\begin{align}
	\label{eqn:12}
	\mathbf{F}^{\left(b\right)'} = \mathbf{F}^{\left(b\right)} + \text{PReLU}\left(\text{GConv1d}\left(\text{LN}\left(\mathbf{F}^{\left(b\right)}\right)\right)\right).
\end{align} 

After that, a Conv1d layer, followed by a SiLU activation function~{\cite{ramachandran2017searching}}, is used to squeeze the channel size to $C^{'}$, which is four times smaller than $C$. Then a band-mixer is adopted for global sub-band shuffling, which is instantiated via a linear matrix operation. After that, another Conv1D+SiLU is adopted to recover the channel size. The process can be expressed as:
\begin{align}
	\label{eqn:13}
	\mathbf{F}^{\left(b\right)''} &= \mathbf{F}^{\left(b\right)'} + \text{SiLU}(\text{Conv1d}(\text{BandMixer}(\text{SiLU}(\nonumber\\
	&\text{Conv1d}(\text{LN}(\mathbf{F}^{(b)'})))))).
\end{align}

Finally, to fully utilize the correlation between adjacent sub-bands, another group convolution with kernel being 3 is adopted with residual connection:
\begin{equation}
	\label{eqn:14}
	\mathbf{F}^{\left(b+1\right)} = \mathbf{F}^{\left(b\right)''} + \text{PReLU}\left(\text{GConv1d}\left(\text{LN}\left(\mathbf{F}^{\left(b\right)''}\right)\right)\right).
\end{equation}
\begin{figure*}[t]
	\centerline{\includegraphics[width=0.94\textwidth]{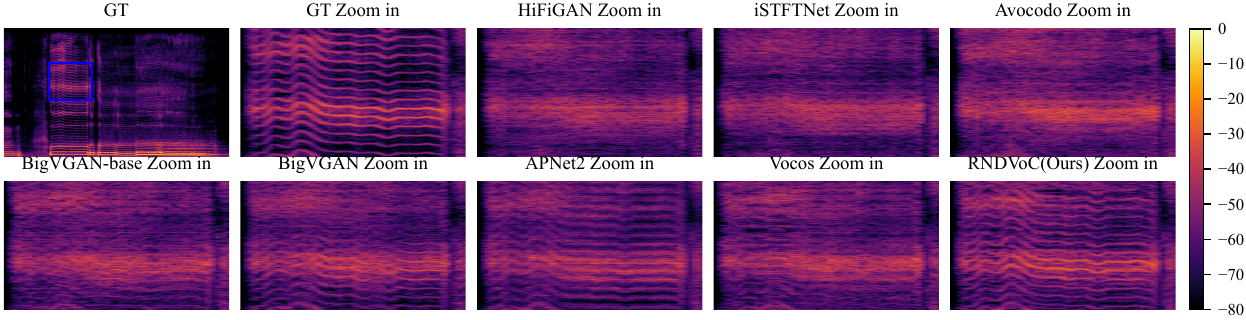}}
	\caption{Spectral visualization of different vocoder methods. The audio clip is a singing voice from the MUSDB18 test set.}
	\label{fig:figure-sota-comparisons}
\end{figure*}

For the narrow-band module (NBM), considering the success of ConvNext v2 blocks in neural vocoder task~{\cite{siuzdakvocos}}, we stack $P=2$ ConvNext v2 blocks to gradually capture long-term relations among adjacent frames. Different from the full-band modeling in previous works, here all sub-bands share the network parameters. Besides, to decrease the computation cost, the number of channels in both hidden and input layers remains unchanged, \emph{i.e.}, $C$. 
\subsection{Loss Function}\label{sec:loss-function}
Following the settings in~{\cite{ai2023apnet,du2023apnet2}}, we adopt both the reconstruction loss and adversarial loss for vocoder training. The reconstruction loss consists of log-amplitude loss $\mathcal{L}_{a}$, phase loss $\mathcal{L}_{p}$, real-imaginary loss $\mathcal{L}_{ri}$, mel-loss $\mathcal{L}_{mel}$, and consistency loss $\mathcal{L}_{c}$:
\begin{align}
	\label{eqn:15}
	\mathcal{L}_{rec} = \lambda_{a}\mathcal{L}_{a} + \lambda_{p}\mathcal{L}_{p} + \lambda_{ri}\mathcal{L}_{ri} + \lambda_{mel}\mathcal{L}_{mel} + \lambda_{c}\mathcal{L}_{c},
\end{align}
where $\left\{\lambda_{a},\lambda_{p},\lambda_{ri},\lambda_{mel},\lambda_{c}\right\}$ are the corresponding hyperparameters. In~{\cite{ai2023apnet,du2023apnet2}}, an anti-wrapping loss is devised, and instantaneous phase (IP), group delay (GD) and instantaneous frequency (IF) are extracted via a specially-designed sparse matrix{\footnote{https://github.com/YangAi520/APCodec/blob/main/models.py}}. However, we find such calculation formula is usually time-consuming as most of the elements in the sparse matrix are zero and have no impact on the multiplied result. Besides, only two directions are considered for phase differential operation, resulting in limited phase relation capture. To this end, we propose a novel omnidirectional phase loss, as shown in Figure~{\ref{fig:omni_loss}}. Specifically, we elaborately design nine $3\times 3$ kernels with fixed parameters $\mathcal{K}=Cat\left(\mathcal{K}_{1},\cdots,\mathcal{K}_{9}\right)\in\mathbb{R}^{9\times 3\times 3}$ to traverse the differential relations with adjacent eight T-F bins, and the fifth kernel is to return the IP. Therefore, with a simple convolution operation, the phase differential can be efficiently implemented as:
\begin{align}
	\label{eqn:16}
	\Delta\mathbf{\Phi} = \mathbf{\Phi}*\mathcal{K}, \Delta\mathbf{\tilde{\Phi}} = \mathbf{\tilde{\Phi}}*\mathcal{K},
\end{align}
where $\left\{\Delta\mathbf{\Phi}, \Delta\mathbf{\tilde{\Phi}}\right\}\in\mathbb{R}^{9\times F\times T}$, and ``$*$'' denotes the convolution operation . Then we can calculate the phase loss with the similar anti-wrapping loss to~{\cite{du2023apnet2}}.

For the adversarial loss, the multi-period discriminator (MPD)~{\cite{kong2020hifi}} and multi-resolution spectrogram discriminator (MRSD)~{\cite{jang21_interspeech}} are utilized. The hinge GAN is adopted, and the adversarial loss for the discriminator can be expressed as:
\begin{align}
	\label{eqn:17}
	\mathcal{L}_D = \frac{1}{M}\sum_{m=1}^{M}\max\left(0,1-D_{m}\left(\mathbf{s}\right)\right) + \max\left(0, 1+D_{m}\left(\mathbf{\tilde{s}}\right)\right),
\end{align}
where $D_{m}$ is the $m$-th discriminator. For the generator, the adversarial loss is:
\begin{equation}
	\label{eqn:18}
	\mathcal{L}_{g} = \frac{1}{M}\sum_{m=1}^{M}\max\left(0, 1 - D_{m}\left(\mathbf{\tilde{s}}\right)\right).
\end{equation}

Besides, the feature matching loss is also utilized, given by:
\begin{equation}
	\label{eqn:19}
	\mathcal{L}_{fm} = \frac{1}{LM}\sum_{l,m}\left|\mathbf{f}_{l}^{m}\left(\mathbf{\tilde{s}}\right) - \mathbf{f}_{l}^{m}\left(\mathbf{s}\right) \right|,
\end{equation}
where $\mathbf{f}_{l}^{m}\left(\cdot\right)$ denotes the $l$-th layer feature for $m$-th sub-discriminator.  The final loss for generator is presented as:
\begin{equation}
	\label{eqn:20}
	\mathcal{L}_{G} = \mathcal{L}_{rec} + \lambda_{g}\mathcal{L}_{g} + \lambda_{fm}\mathcal{L}_{fm},
\end{equation}
where $\left\{\lambda_{g}, \lambda_{fm}\right\}$ are the corresponding hyperparameters. Detailed settings can be found in the supplementary material. 

\renewcommand\arraystretch{1.20}
\begin{table}[t]
	\centering
	\Huge
	\resizebox{0.47\textwidth}{!}{
		\begin{tabular}{c|ccccc}
			\toprule
			\multirow{2}*{Models} &\multirow{2}*{PESQ$\uparrow$} &Periodicity$\downarrow$ &V/UV$\uparrow$ &Pitch$\downarrow$ &\multirow{2}*{VISQOL$\uparrow$}\\
			& &RMSE &F1 &RMSE & \\
			\midrule
			WaveGlow-256$^{\dagger}$  &3.138  &0.1485 &0.9378 &- &-\\
			HiFiGAN-V1  &3.056  &0.1671 &0.9212 &52.5285 &4.7209 \\
			iSTFTNet-V1  &2.880 &0.1672 &0.9177 &53.0724 &4.6548 \\
			UnivNet-c32$\dagger$   &3.277 &0.1305 &0.9347 &41.5110 &4.7530\\
			Avocodo    &3.217 &0.1611 &0.9134 &51.5998 &4.7620\\
			BigVGAN-base(1M steps)$\dagger$   &3.519 &0.1287 &0.9459 &- &- \\
			BigVGAN(1M steps)$\dagger$   &4.027 &0.1018 &0.9598 &-
			&-\\
			BigVGAN-base(5M steps)$\dagger$   &3.841 &0.1073 &0.9540 &32.5413 &4.9067 \\
			BigVGAN(5M steps)$\dagger$  &\textbf{4.269} &\underline{0.0790} &\underline{0.9670} &\underline{24.2814}
			&\textbf{4.9632}\\
			APNet    &2.897 &0.1586 &0.9265 &39.6629 &4.6659\\
			APNet2  &2.834 &0.1529 &0.9227 &46.3732 &4.5817 \\
			Vocos$\dagger$   &3.615 &0.1146 &0.9484 &35.5844 &4.8785\\
			\hline
			\textbf{RNDVoC(Ours)}  &\underline{4.226} &\textbf{0.0742} &\textbf{0.9698} &\textbf{23.9658} &\underline{4.9154}\\
			\hline
	\end{tabular}}
	\caption{Objective comparisons among baselines on the LibriTTS benchmark. ``-'' denotes the results are not reported, and $\dagger$ denotes the results are calculated using the open-sourced model checkpoints.}
	\label{tbl:objective-metric-libritts}
\end{table}
\begin{table*}[ht]
	\centering
	\Huge
	\resizebox{0.95\textwidth}{!}{
		\begin{tabular}{ccccccccc}
			\toprule
			Models &GT &HiFiGAN-V1 &Avocodo &BigVGAN-base &BigVGAN &APNet2 &Vocos &RNDVoC\\
			MUSHRA &89.45$\pm$0.45 &69.99$\pm$1.15 &68.16$\pm$1.26 &74.27$\pm$1.12 &79.33$\pm$0.92 &54.95$\pm$1.11 &73.18$\pm$1.15 &\textbf{**80.74$\pm$0.99}\\
			\bottomrule
			\hline \multicolumn{8}{l}{Note: $^{**}p<0.05$, $^{*}p<0.1$} 
	\end{tabular}}
	\caption{MUSHRA scores among different methods on the LibriTTS benchmark. The confidence level is 95\%, and we performed a t-test comparing RNDVoC with BigVGAN.}
	\label{tbl:libritts-mushra}
\end{table*}
\section{Experiments}\label{sec:experiments}
\subsection{Datasets}\label{sec:datasets}
Two benchmarks are employed in this study, namely LJSpeech~{\cite{ljspeech17}} and LibriTTS~{\cite{zen19_interspeech}}. The LJSpeech dataset includes 13,100 clean speech clips by a single female, and the sampling rate is 22.05 kHz. Following the division in the open-sourced VITS repository{\footnote{https://github.com/jaywalnut310/vits/tree/main/filelists}}, $\left\{12500,100,500\right\}$ clips are used for training, valiation, and testing, respectively. The LibriTTS dataset covers diverse recording environments with the sampling rate of 24 kHz. Following the division in~{\cite{leebigvgan}}, \{\textit{train-clean-100}, \textit{train-clean-300}, \textit{train-other-500}\} are for model training. The subsets $\textit{dev-clean}+\textit{dev-other}$ are for objective comparisons, and $\textit{test-clean}+\textit{test-other}$ are for subjective evaluations. 
\subsection{Configurations}\label{sec:mel-configurations}
For the LJSpeech dataset, the number of mel-spectrogram $F_{m}$ is set to 80, and the upper-bound frequency $f_{max}$ is 8 kHz. For the LibriTTS dataset, 100 mel-bands are used with $f_{\max} = 12$ kHz. For both benchmarks, 1024-point FFT is set, with 1024 Hann window and 256 hop size.

A batch size of 16, a segment size of 16384, and an initial learning rate of 2e-4 are used for training. The AdamW optimizer~{\cite{loshchilov2017decoupled}} is employed, with $\left\{\beta_{1}=0.8, \beta_{2}=0.99\right\}$. The generator and discriminator are updated for 1 million steps, respectively.
\renewcommand\arraystretch{1.20}
\begin{table}
	\centering
	\Huge
	\resizebox{0.49\textwidth}{!}{
		\begin{tabular}{|cc|ccccc|}
			\hline
			\multirow{2}*{Ids} &\multicolumn{1}{c|}{\multirow{2}*{Setting}} &\multirow{2}*{PESQ$\uparrow$} &\multirow{2}*{MCD$\downarrow$} &Periodicity$\downarrow$ &V/UV$\uparrow$ &Pitch$\downarrow$\\
			& & & &RMSE &F1 &RMSE\\
			\hline
			1 &Baseline &3.987 &2.0471 &0.0854 &0.9714 &21.3183\\
			\hline 
			2 &Remove omnidirectional phase loss &3.892 &2.2137 &0.0889 &0.9697 &21.9124 \\
			3 &\indent Remove RND mode &3.655 &2.4976 &0.9611 &0.1114 &26.0120 \\
			4 &\indent{Set matrice} $\left\{\mathcal{A}, \mathcal{A}^{\dagger}\right\}$ as learnable &3.645 &2.5989 &0.9591 &0.1164 &25.6292 \\
			\hline
	\end{tabular}}
	\caption{Ablation studies conducted on the LJSpeech dataset.}
	\label{tbl:ablations}
\end{table}
\begin{figure}
	\centering
	\includegraphics[width=0.46\textwidth]{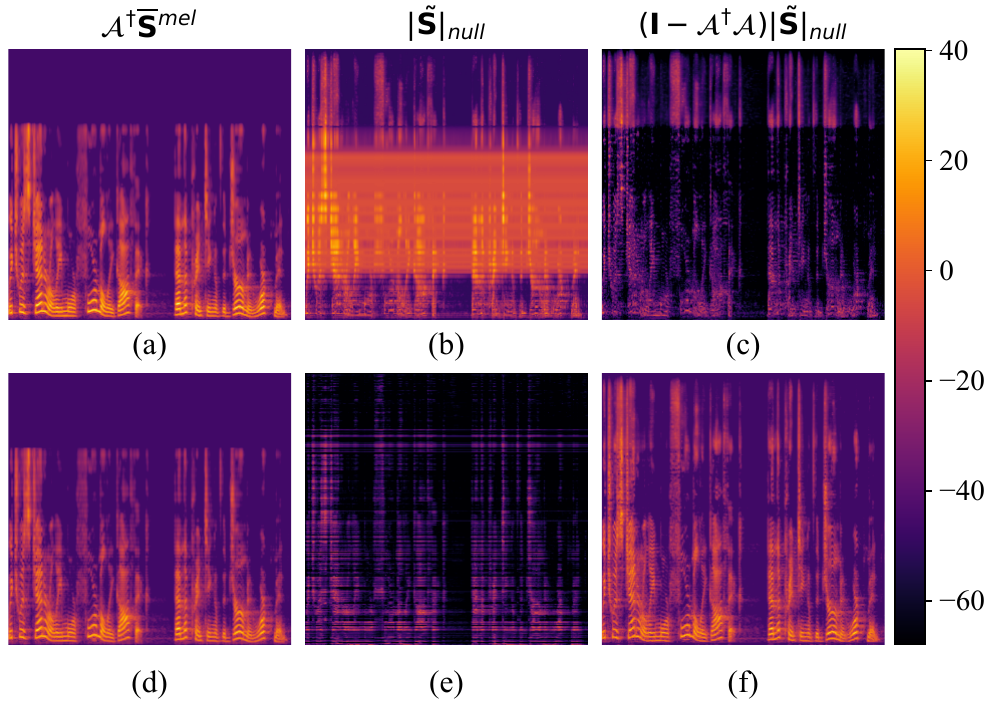}
	\caption{Spectral visualization of the range-space and nullspace with respect to whether $\left\{\mathcal{A}^{\dagger}, \mathcal{A}\right\}$ are fixed. (a)-(c) Thematrix parameters are fixed. (d)-(f) The matrix parametersare learnable.}  
	\label{fig:rnd-visualization}
\end{figure}
\subsection{Results and Analysis}\label{sec:results-and-analysis}
In this study, various representative time-domain and T-F domain-based baselines are chosen for comparisons, including HiFiGAN~{\cite{kong2020hifi}}{\footnote{https://github.com/jik876/hifi-gan}}, iSTFTNet~{\cite{kaneko2022istftnet}}{\footnote{https://github.com/rishikksh20/iSTFTNet-pytorch}}, Avocodo~{\cite{bak2023avocodo}}{\footnote{https://github.com/ncsoft/avocodo}}, WaveGlow~{\cite{prenger2019waveglow}}{\footnote{https://github.com/NVIDIA/waveglow}}, UnivNet~{\cite{jang21_interspeech}}{\footnote{https://github.com/maum-ai/univnet}}, BigVGAN~{\cite{leebigvgan}}{\footnote{https://github.com/NVIDIA/BigVGAN}}, APNet~{\cite{ai2023apnet}}, APNet2~{\cite{du2023apnet2}}{\footnote{https://github.com/redmist328/APNet2}}, FreeV~{\cite{lv24_interspeech}}{\footnote{https://github.com/BakerBunker/FreeV}}, and Vocos~{\cite{siuzdakvocos}}{\footnote{https://github.com/gemelo-ai/vocos}}. 

Five metrics are involved in the objective evaluations: (1) Multi-resolution STFT (M-STFT)~{\cite{yamamoto2020parallel}} is used to evaluate the spectral distance across multiple resolutions. (2) Wide-band version of Perceptual evaluation of speech quality (PESQ)~{\cite{rec2005p}} is chosen to assess the objective speech quality. (3) Mel-cepstral distortion (MCD)~{\cite{kubichek1993mel}} measures the difference between mel-spectrograms through dynamic time wrapping (DTW). (4) Periodicity RMSE, V/UV F1 score, and pitch RMSE~{\cite{morrisonchunked}}, which are regarded as major artifacts for non-autoregressive neural vocoders. (5) Virtual Speech Quality Objective Listener (VISQOL)~{\cite{hines2015visqol}} predicts the Mean Opinion Score-Listening Quality Objective (MOS-LQO) score by evaluating the spectro-temporal similarity.

Except for objective metrics, we also conduct the MUSHRA testing based on the BeaqleJS platform~{\cite{kraft2014beaqlejs}} for subjective evaluations. Thirty-five participants majoring in audio signal processing are engaged in the testing. For the former, each person rates the speech processed by different algorithms on a scale ranging from 0 to 100 in terms of the overall similarity compared to the reference clips.
\begin{table}
	\centering
	\Huge
	\resizebox{0.49\textwidth}{!}{
		\begin{tabular}{ccc|c|cc}
			\toprule
			\multirow{2}*{Models} &\#Parm. &\#MACs &\multirow{2}*{Datasets} 
			&\multirow{2}*{PESQ$\uparrow$}  &\multirow{2}*{VISQOL$\uparrow$} \\
			&(M) &(Giga/5s) & & & \\
			\midrule
			\multirow{2}*{HiFiGAN-V2} &\multirow{2}*{0.92} &9.6  &LJSpeech  &2.848 &4.5419\\
			& &10.46 &LibriTTS  &2.308 &4.3695 \\
			\midrule
			\multirow{2}*{RNDVoC-Lite} &\multirow{2}*{0.71} &9.54 &LJSpeech &3.769 &4.7817 \\
			& &10.39 &LibriTTS &3.834 &4.8736\\
			\midrule
			\multirow{2}*{RNDVoC-UltraLite} &\multirow{2}*{0.08} &1.66 &LJSpeech  &3.264 &4.7006 \\
			& &1.81 &LibriTTS &3.499 &4.8044\\
			\bottomrule
	\end{tabular}}
	\caption{Objective comparisons among lightweight neural vocoders.}
	\label{tbl:light-weight}
\end{table}
\subsubsection{Comparisons with SOTA Methods}\label{comparisons-with-sota}
Tables~{{\ref{tbl:objective-metric-ljs}} and {\ref{tbl:objective-metric-libritts}} present the objective comparisons on the LJSpeech and LibriTTS datasets, respectively. Several valuable observations can be made. First, T-F domain-based methods exhibit overall faster inference speed over time-domain-based methods. This is mainly due to the use of STFT and its inverse transform, \emph{i.e.}, iSTFT, and no upsampling opeartion is required. Second, T-F domain based methods possess overall notably less computation complexity, \emph{e.g.}, 5.8 GMACs of Vocos vs. 152.9 GMACs of HiFiGAN. This has made T-F domain based neural vocoders attractive more recently. Third, compared with BigVGAN, the speech quality of existing T-F domain based neural vocoders are still notably inferior. Thanks to the fined-grained modeling along the time and sub-band axes, the proposed RNDVoC enjoys both light-weight network structure and promising performance. To be specific, with less than 3\% trainable parameters and 10\% computation cost, our approach yields comparable performance over BigVGAN in the LJSpeech dataset, and better performance in the LibriTTS benchmark. It is noteworthy that even compared with BigVGAN trained for 5M steps, our method still behaves competitive. It fully validates the effectiveness of the proposed approach.
	
Table~{\ref{tbl:libritts-mushra}} gives the MUSHRA results on the test set of LibriTTS dataset. One can observe that our RNDVoC is superior to BigVGAN with statistical difference ($p < 0.05$), further validating the advantage of our method in subjective quality. We notice that the subjective score of APNet2 is notably inferior to other methods. According to the feedback from the listeners, audible husky buzzing artifacts can arise in the generated utterances from APNet2, which leads to biased low score.

Figure~{\ref{fig:figure-sota-comparisons}} shows the spectral visualizations among different models, where the clip is a vocal voice from the out-of-distribution MUSDB18~{\cite{rafii2017musdb18}} test set. Evidently, compared with other baselines, our approach can better recover the harmonic details.
\begin{figure}
	\centering
	\includegraphics[width=0.46\textwidth]{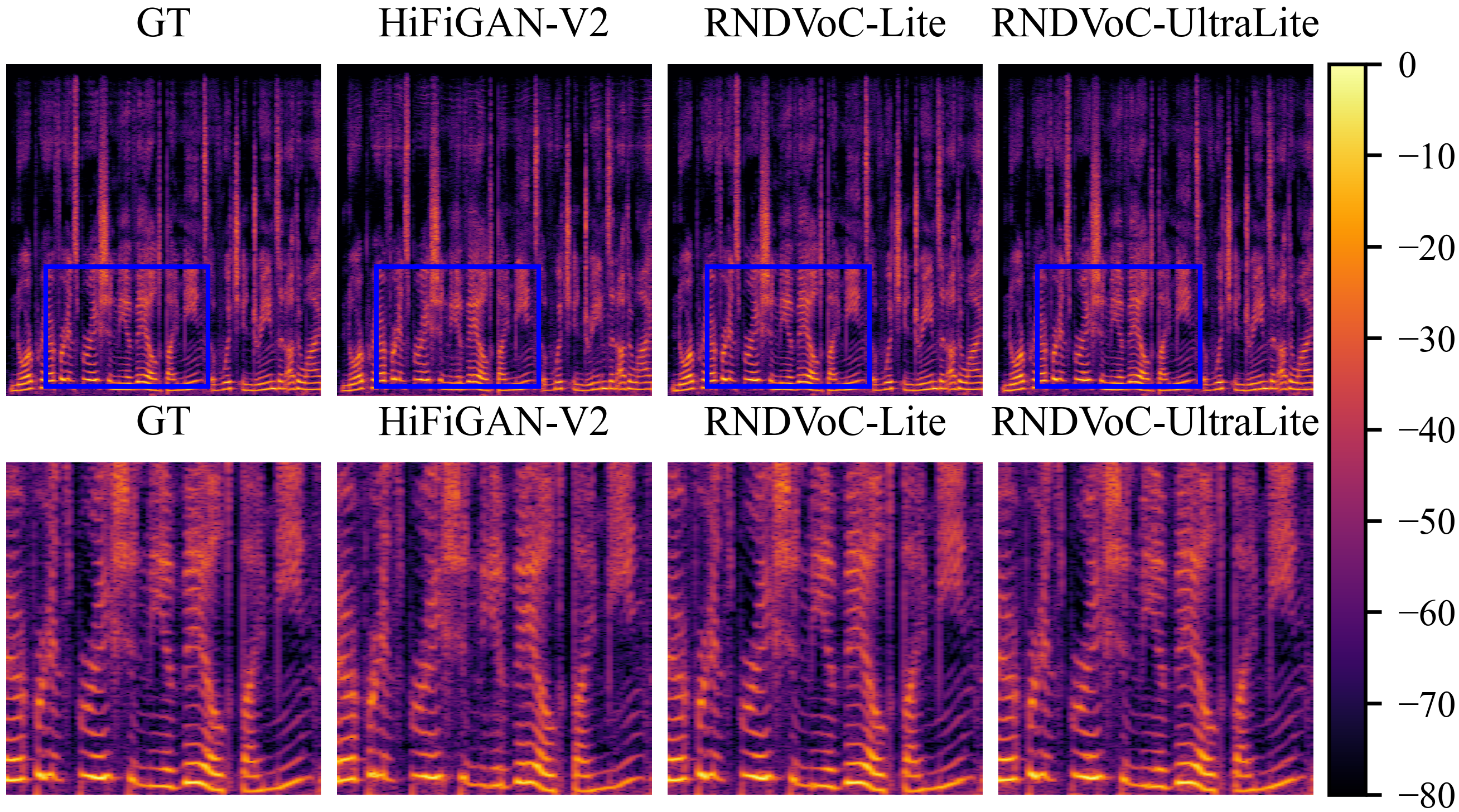}
	\caption{Spectral visualization generated by three light-weight neural vocoders, namely HiFiGAN-V2, RNDVoC-Lite, and RNDVoC-UltraLite. The audio clip is a speech voice from the LibriTTS test set.}  
	\label{fig:lightweight-visualization}
\end{figure}
\subsubsection{Ablation Studies}\label{sec:ablations}
Table~{\ref{tbl:ablations}} presents the ablation studies based on the LJSpeech dataset. First, we replace the proposed omnidirectional phase loss with that in~{\cite{du2023apnet2}}, as shown from id1 to id2. One can observe clear performance degradation in objective metrics, indicating the effectiveness of the omnidirectional phase loss. Based on id2, we further remove the RND mode in target reconstruction, \emph{i.e.}, the explicit superimposition between $|\tilde{\mathbf{S}}|_{range}$ and $\left(\mathbf{I} - \mathcal{A}^{\dagger}\mathcal{A}\right)|\tilde{\mathbf{S}}|_{null}$ is changed into the direct target mapping. From id2 to id3, notable performance degradation can be observed, indicating the significance of the proposed RND modeling. Then based on id2, we set the matrice $\left\{\mathcal{A}^{\dagger}, \mathcal{A}\right\}$ as learnable, trying to break the orthogonality between the two sub-spaces. From id2 to id4, we also observe the performance degradation, indicating the significance of orthogonality between two sub-spaces. In Figure~{\ref{fig:rnd-visualization}}, we visualize the estimated components form the range-space and null-space in terms of whether $\left\{\mathcal{A}^{\dagger}, \mathcal{A}\right\}$ are fixed. One can observe that if the parameters are fixed, the null-space, \emph{i.e.}, $\left(\mathbf{I} - \mathcal{A}^{\dagger}\mathcal{A}\right)|\tilde{\mathbf{S}}|_{null}$ can capture the sparse and detailed harmonic structure, as shown in Figure~{\ref{fig:rnd-visualization}}(c). In contrast, in the learnable scenario, the estimation of the null-space is no longer sparse, indicating that the orthogonality property may not hold.
\subsubsection{Toward Light-weight Design}\label{toward-light-weight-design} 
To meet the requirements of edge-devices in practical scenarios, we investigate the potential of our method in light-weight design. Concretely, the number of DPB $B$ is reduced into 4. When the channel size $C$ is squeezed to 128 and 32, we obtain the light and ultra-lite version, namely abbreviated as \textbf{RNDVoC-Lite} and \textbf{RNDVoC-UltraLite}. Table~{\ref{tbl:light-weight}} presents the performance of different light-weight models. Evidently, thanks to the efficacy of RND in information preservation, where the learning network only needs to concentrate on the null-space part, even only with 0.08M, our method still notably outperforms HiFiGAN-V2 in PESQ and VISQOL, which validates the effectiveness of our method.

In Figure~{\ref{fig:lightweight-visualization}}, we present the spectral representations reconstructed by three lightweight vocoders. Notably, while HiFiGAN-V2 exhibits significant harmonic blurring (see the blue-boxed zoomed region), both RNDVoC-Lite and RNDVoC-UltraLite retain clear harmonic details. This contrast further highlights the superiority of our approach in lightweight design scenarios.

\section{Conclusion}\label{sec:conclusion}
In this paper, we propose an innovative T-F domain-based neural vocoders. We bridge the connection with classical range-null decomposition theory, where the range-space aims to convert the original acoustic feature in the mel-scale domain into the target linear-scale domain, and null-space serves to reconstruct the remaining spectral details. Based on that, a novel dual-path network structure is devised, where the cross-band and narrow-band modules are devised to efficiently model the relations in the sub-band and time axes. Extensive experiments are conducted on the LJSpeech and LibriTTS benchmarks to validate the efficacy of the proposed method. 

\bibliographystyle{named}
\bibliography{ijcai25}

\begin{thebibliography}{}

\bibitem[\protect\citeauthoryear{Ai and Ling}{2023}]{ai2023apnet}
Yang Ai and Zhen-Hua Ling.
\newblock {APNet}: An all-frame-level neural vocoder incorporating direct prediction of amplitude and phase spectra.
\newblock {\em IEEE/ACM Trans. Audio, Speech, Lang. Process.}, 31:2145--2157, 2023.

\bibitem[\protect\citeauthoryear{Bak \bgroup \em et al.\egroup }{2023}]{bak2023avocodo}
Taejun Bak, Junmo Lee, Hanbin Bae, Jinhyeok Yang, Jae-Sung Bae, and Young-Sun Joo.
\newblock Avocodo: Generative adversarial network for artifact-free vocoder.
\newblock In {\em Proc. AAAI}, volume~37, pages 12562--12570, 2023.

\bibitem[\protect\citeauthoryear{Baraniuk \bgroup \em et al.\egroup }{2010}]{baraniuk2010model}
Richard~G Baraniuk, Volkan Cevher, Marco~F Duarte, and Chinmay Hegde.
\newblock Model-based compressive sensing.
\newblock {\em IEEE Trans. Inf. Theory}, 56(4):1982--2001, 2010.

\bibitem[\protect\citeauthoryear{Chen \bgroup \em et al.\egroup }{2021}]{chenwavegrad}
Nanxin Chen, Yu~Zhang, Heiga Zen, Ron~J Weiss, Mohammad Norouzi, and William Chan.
\newblock Wave{G}rad: Estimating {G}radients for {W}aveform {G}eneration.
\newblock In {\em Proc. ICLR}, 2021.

\bibitem[\protect\citeauthoryear{Dinh \bgroup \em et al.\egroup }{2016}]{dinh2016density}
Laurent Dinh, Jascha Sohl-Dickstein, and Samy Bengio.
\newblock Density estimation using real nvp.
\newblock {\em arXiv preprint arXiv:1605.08803}, 2016.

\bibitem[\protect\citeauthoryear{Du \bgroup \em et al.\egroup }{2023}]{du2023apnet2}
Hui-Peng Du, Ye-Xin Lu, Yang Ai, and Zhen-Hua Ling.
\newblock {APNet2}: High-{Q}uality and {H}igh-{E}fficiency {N}eural {V}ocoder with {D}irect {P}rediction of {A}mplitude and {P}hase {S}pectra.
\newblock In {\em Proc. NCMSC}, pages 66--80. Springer, 2023.

\bibitem[\protect\citeauthoryear{He \bgroup \em et al.\egroup }{2015}]{he2015delving}
Kaiming He, Xiangyu Zhang, Shaoqing Ren, and Jian Sun.
\newblock Delving {D}eep into {R}ectifiers: Surpassing {H}uman-{L}evel {P}erformance on {I}magenet {C}lassification.
\newblock In {\em Proc. ICCV}, pages 1026--1034, 2015.

\bibitem[\protect\citeauthoryear{Hines \bgroup \em et al.\egroup }{2015}]{hines2015visqol}
Andrew Hines, Jan Skoglund, Anil~C Kokaram, and Naomi Harte.
\newblock {ViSQOL}: an objective speech quality model.
\newblock {\em EURASIP J. Audio Speech Music Process.}, 2015:1--18, 2015.

\bibitem[\protect\citeauthoryear{Huang \bgroup \em et al.\egroup }{2022a}]{huang2022fastdiff}
Rongjie Huang, Max~WY Lam, Jun Wang, Dan Su, Dong Yu, Yi~Ren, and Zhou Zhao.
\newblock Fast{D}iff: A {F}ast {C}onditional {D}iffusion {M}odel for {H}igh-{Q}uality {S}peech {S}ynthesis.
\newblock In {\em Proc.~{IJCAI}}, pages 4157--4163, 2022.

\bibitem[\protect\citeauthoryear{Huang \bgroup \em et al.\egroup }{2022b}]{huang2022prodiff}
Rongjie Huang, Zhou Zhao, Huadai Liu, Jinglin Liu, Chenye Cui, and Yi~Ren.
\newblock Prodiff: Progressive fast diffusion model for high-quality text-to-speech.
\newblock In {\em Proc. ACMMM}, pages 2595--2605, 2022.

\bibitem[\protect\citeauthoryear{Jang \bgroup \em et al.\egroup }{2021}]{jang21_interspeech}
Won Jang, Dan Lim, Jaesam Yoon, Bongwan Kim, and Juntae Kim.
\newblock Univ{N}et: A neural {V}ocoder with {M}ulti-resolution {S}pectrogram {D}iscriminators for {H}igh-{F}idelity {W}aveform {G}eneration.
\newblock In {\em Proc. Interspeech}, pages 2207--2211, 2021.

\bibitem[\protect\citeauthoryear{Juvela \bgroup \em et al.\egroup }{2019}]{juvela2019glotnet}
Lauri Juvela, Bajibabu Bollepalli, Vassilis Tsiaras, and Paavo Alku.
\newblock Glotnet—a raw waveform model for the glottal excitation in statistical parametric speech synthesis.
\newblock {\em IEEE/ACM Trans. Audio, Speech, Lang. Process.}, 27(6):1019--1030, 2019.

\bibitem[\protect\citeauthoryear{Kalchbrenner \bgroup \em et al.\egroup }{2018}]{kalchbrenner2018efficient}
Nal Kalchbrenner, Erich Elsen, Karen Simonyan, Seb Noury, Norman Casagrande, Edward Lockhart, Florian Stimberg, Aaron Oord, Sander Dieleman, and Koray Kavukcuoglu.
\newblock Efficient neural audio synthesis.
\newblock In {\em Proc. ICML}, pages 2410--2419. PMLR, 2018.

\bibitem[\protect\citeauthoryear{Kaneko \bgroup \em et al.\egroup }{2022}]{kaneko2022istftnet}
Takuhiro Kaneko, Kou Tanaka, Hirokazu Kameoka, and Shogo Seki.
\newblock i{STFTN}et: Fast and lightweight mel-spectrogram vocoder incorporating inverse short-time fourier transform.
\newblock In {\em Proc. ICASSP}, pages 6207--6211. IEEE, 2022.

\bibitem[\protect\citeauthoryear{Kawahara}{2006}]{kawahara2006straight}
Hideki Kawahara.
\newblock {STRAIGHT}, exploitation of the other aspect of {VOCODER}: Perceptually isomorphic decomposition of speech sounds.
\newblock {\em Acoustical science and technology}, 27(6):349--353, 2006.

\bibitem[\protect\citeauthoryear{Keith and Linda}{2017}]{ljspeech17}
I.~Keith and J.~Linda.
\newblock The {LJ} {S}peech {D}ataset.
\newblock \url{https://keithito.com/LJ-Speech-Dataset/}, 2017.
\newblock Accessed: 2025-01-12.

\bibitem[\protect\citeauthoryear{Kim \bgroup \em et al.\egroup }{2019}]{kim2019flowavenet}
Sungwon Kim, Sang-gil Lee, Jongyoon Song, Jaehyeon Kim, and Sungroh Yoon.
\newblock Flo{W}ave{N}et: A {G}enerative {F}low for {R}aw {A}udio.
\newblock In {\em Proc. ICML}, pages 3370--3378. PMLR, 2019.

\bibitem[\protect\citeauthoryear{Kong \bgroup \em et al.\egroup }{2020}]{kong2020hifi}
Jungil Kong, Jaehyeon Kim, and Jaekyoung Bae.
\newblock {Hifi-GAN}: Generative adversarial networks for efficient and high fidelity speech synthesis.
\newblock In {\em Proc. NeurIPS}, pages 17022--17033, 2020.

\bibitem[\protect\citeauthoryear{Kong \bgroup \em et al.\egroup }{2021}]{kongdiffwave}
Zhifeng Kong, Wei Ping, Jiaji Huang, Kexin Zhao, and Bryan Catanzaro.
\newblock Diff{W}ave: A {V}ersatile {D}iffusion {M}odel for {A}udio {S}ynthesis.
\newblock In {\em Proc. ICLR}, 2021.

\bibitem[\protect\citeauthoryear{Kraft and Z{\"o}lzer}{2014}]{kraft2014beaqlejs}
Sebastian Kraft and Udo Z{\"o}lzer.
\newblock Beaqle{JS}: {HTML5} and {JavaScript} based framework for the subjective evaluation of audio quality.
\newblock In {\em Linux Audio Conference, Karlsruhe, DE}, 2014.

\bibitem[\protect\citeauthoryear{Kubichek}{1993}]{kubichek1993mel}
Robert Kubichek.
\newblock Mel-cepstral distance measure for objective speech quality assessment.
\newblock In {\em Proceedings of IEEE pacific rim conference on communications computers and signal processing}, volume~1, pages 125--128. IEEE, 1993.

\bibitem[\protect\citeauthoryear{Kumar \bgroup \em et al.\egroup }{2019}]{kumar2019melgan}
Kundan Kumar, Rithesh Kumar, Thibault De~Boissiere, Lucas Gestin, Wei~Zhen Teoh, Jose Sotelo, Alexandre De~Brebisson, Yoshua Bengio, and Aaron~C Courville.
\newblock Melgan: Generative adversarial networks for conditional waveform synthesis.
\newblock {\em Proc. NeurIPS}, 32, 2019.

\bibitem[\protect\citeauthoryear{Lee \bgroup \em et al.\egroup }{2022}]{leepriorgrad}
Sang-gil Lee, Heeseung Kim, Chaehun Shin, Xu~Tan, Chang Liu, Qi~Meng, Tao Qin, Wei Chen, Sungroh Yoon, and Tie-Yan Liu.
\newblock Prior{G}rad: Improving {C}onditional {D}enoising {D}iffusion {M}odels with {D}ata-{D}ependent {A}daptive {P}rior.
\newblock In {\em Proc. ICLR}, 2022.

\bibitem[\protect\citeauthoryear{Lee \bgroup \em et al.\egroup }{2023}]{leebigvgan}
Sang-gil Lee, Wei Ping, Boris Ginsburg, Bryan Catanzaro, and Sungroh Yoon.
\newblock {BigVGAN}: A {U}niversal {N}eural {V}ocoder with {L}arge-{S}cale {T}raining.
\newblock In {\em Proc. ICLR}, 2023.

\bibitem[\protect\citeauthoryear{Lei~Ba \bgroup \em et al.\egroup }{2016}]{lei2016layer}
Jimmy Lei~Ba, Jamie~Ryan Kiros, and Geoffrey~E Hinton.
\newblock Layer normalization.
\newblock {\em ArXiv e-prints}, pages arXiv--1607, 2016.

\bibitem[\protect\citeauthoryear{Li \bgroup \em et al.\egroup }{2022}]{ijcai2022p582}
Andong Li, Shan You, Guochen Yu, Chengshi Zheng, and Xiaodong Li.
\newblock {Taylor, Can You Hear Me Now? A Taylor-Unfolding Framework for Monaural Speech Enhancement}.
\newblock In {\em Proc. IJCAI}, pages 4193--4200, 2022.

\bibitem[\protect\citeauthoryear{Li \bgroup \em et al.\egroup }{2025}]{li2025neural}
Andong Li, Zhihang Sun, Fengyuan Hao, Xiaodong Li, and Chengshi Zheng.
\newblock Neural vocoders as speech enhancers.
\newblock {\em arXiv preprint arXiv:2501.13465}, 2025.

\bibitem[\protect\citeauthoryear{Liu \bgroup \em et al.\egroup }{2023}]{liu2023audioldm}
Haohe Liu, Zehua Chen, Yi~Yuan, Xinhao Mei, Xubo Liu, Danilo Mandic, Wenwu Wang, and Mark~D Plumbley.
\newblock Audio{LDM}: Text-to-{A}udio {G}eneration with {L}atent {D}iffusion {M}odels.
\newblock In {\em Proc. ICML}, pages 21450--21474. PMLR, 2023.

\bibitem[\protect\citeauthoryear{Loshchilov and Hutter}{2017}]{loshchilov2017decoupled}
Ilya Loshchilov and Frank Hutter.
\newblock Decoupled weight decay regularization.
\newblock {\em arXiv preprint arXiv:1711.05101}, 2017.

\bibitem[\protect\citeauthoryear{Lu \bgroup \em et al.\egroup }{2022}]{lu2022dpm}
Cheng Lu, Yuhao Zhou, Fan Bao, Jianfei Chen, Chongxuan Li, and Jun Zhu.
\newblock {DPM-solver}: A fast ode solver for diffusion probabilistic model sampling in around 10 steps.
\newblock {\em Proc. NeurIPS}, 35:5775--5787, 2022.

\bibitem[\protect\citeauthoryear{Lv \bgroup \em et al.\egroup }{2024}]{lv24_interspeech}
Yuanjun Lv, Hai Li, Ying Yan, Junhui Liu, Danming Xie, and Lei Xie.
\newblock Free{V}: Free {L}unch {F}or {V}ocoders {T}hrough {P}seudo {I}nversed {M}el {F}ilter.
\newblock In {\em Proc. Interspeech}, pages 3869--3873, 2024.

\bibitem[\protect\citeauthoryear{Morise \bgroup \em et al.\egroup }{2016}]{morise2016world}
Masanori Morise, Fumiya Yokomori, and Kenji Ozawa.
\newblock World: a vocoder-based high-quality speech synthesis system for real-time applications.
\newblock {\em IEICE Trans Inf Syst}, 99(7):1877--1884, 2016.

\bibitem[\protect\citeauthoryear{Morrison \bgroup \em et al.\egroup }{2022}]{morrisonchunked}
Max Morrison, Rithesh Kumar, Kundan Kumar, Prem Seetharaman, Aaron Courville, and Yoshua Bengio.
\newblock Chunked {A}utoregressive {GAN} for {C}onditional {W}aveform {S}ynthesis.
\newblock In {\em Proc. ICLR}, 2022.

\bibitem[\protect\citeauthoryear{Oord \bgroup \em et al.\egroup }{2018}]{oord2018parallel}
Aaron Oord, Yazhe Li, Igor Babuschkin, Karen Simonyan, Oriol Vinyals, Koray Kavukcuoglu, George Driessche, Edward Lockhart, Luis Cobo, Florian Stimberg, et~al.
\newblock Parallel wavenet: Fast high-fidelity speech synthesis.
\newblock In {\em Proc. ICML}, pages 3918--3926. PMLR, 2018.

\bibitem[\protect\citeauthoryear{Ping \bgroup \em et al.\egroup }{2020}]{ping2020waveflow}
Wei Ping, Kainan Peng, Kexin Zhao, and Zhao Song.
\newblock Waveflow: A compact flow-based model for raw audio.
\newblock In {\em Proc. ICML}, pages 7706--7716. PMLR, 2020.

\bibitem[\protect\citeauthoryear{Prenger \bgroup \em et al.\egroup }{2019}]{prenger2019waveglow}
Ryan Prenger, Rafael Valle, and Bryan Catanzaro.
\newblock Waveglow: A flow-based generative network for speech synthesis.
\newblock In {\em Proc. ICASSP}, pages 3617--3621. IEEE, 2019.

\bibitem[\protect\citeauthoryear{Quan and Li}{2024}]{quan2024spatialnet}
Changsheng Quan and Xiaofei Li.
\newblock Spatial{N}et: Extensively learning spatial information for multichannel joint speech separation, denoising and dereverberation.
\newblock {\em IEEE/ACM Trans. Audio, Speech, Lang. Process.}, 32:1310--1323, 2024.

\bibitem[\protect\citeauthoryear{Rafii \bgroup \em et al.\egroup }{2017}]{rafii2017musdb18}
Zafar Rafii, Antoine Liutkus, Fabian-Robert St{\"o}ter, Stylianos~Ioannis Mimilakis, and Rachel Bittner.
\newblock The {MUSDB18} corpus for music separation.
\newblock 2017.

\bibitem[\protect\citeauthoryear{Ramachandran \bgroup \em et al.\egroup }{2017}]{ramachandran2017searching}
Prajit Ramachandran, Barret Zoph, and Quoc~V Le.
\newblock Searching for activation functions.
\newblock {\em arXiv preprint arXiv:1710.05941}, 2017.

\bibitem[\protect\citeauthoryear{Rec}{2005}]{rec2005p}
ITUT Rec.
\newblock P. 862.2: Wideband extension to recommendation p. 862 for the assessment of wideband telephone networks and speech codecs.
\newblock {\em International Telecommunication Union, CH--Geneva}, 41:48--60, 2005.

\bibitem[\protect\citeauthoryear{Siuzdak}{2024}]{siuzdakvocos}
Hubert Siuzdak.
\newblock Vocos: Closing the gap between time-domain and fourier-based neural vocoders for high-quality audio synthesis.
\newblock In {\em Proc. ICLR}, 2024.

\bibitem[\protect\citeauthoryear{Valin and Skoglund}{2019}]{valin2019lpcnet}
Jean-Marc Valin and Jan Skoglund.
\newblock L{PCN}et: Improving neural speech synthesis through linear prediction.
\newblock In {\em Proc. ICASSP}, pages 5891--5895. IEEE, 2019.

\bibitem[\protect\citeauthoryear{Van Den~Oord \bgroup \em et al.\egroup }{2016}]{van2016wavenet}
Aaron Van Den~Oord, Sander Dieleman, Heiga Zen, Karen Simonyan, Oriol Vinyals, Alex Graves, Nal Kalchbrenner, Andrew Senior, Koray Kavukcuoglu, et~al.
\newblock Wavenet: A generative model for raw audio.
\newblock {\em arXiv preprint arXiv:1609.03499}, 12, 2016.

\bibitem[\protect\citeauthoryear{Wang \bgroup \em et al.\egroup }{2017}]{wang2017tacotron}
Yuxuan Wang, RJ~Skerry-Ryan, Daisy Stanton, Yonghui Wu, Ron~J Weiss, Navdeep Jaitly, Zongheng Yang, Ying Xiao, Zhifeng Chen, Samy Bengio, et~al.
\newblock Tacotron: Towards {E}nd-to-{E}nd {S}peech {S}ynthesis.
\newblock In {\em Proc. {I}nterspeech}, pages 4006--4010, 2017.

\bibitem[\protect\citeauthoryear{Woo \bgroup \em et al.\egroup }{2023}]{woo2023convnext}
Sanghyun Woo, Shoubhik Debnath, Ronghang Hu, Xinlei Chen, Zhuang Liu, In~So Kweon, and Saining Xie.
\newblock Convnext v2: Co-designing and scaling convnets with masked autoencoders.
\newblock In {\em Proc. CVPR}, pages 16133--16142, 2023.

\bibitem[\protect\citeauthoryear{Yamamoto \bgroup \em et al.\egroup }{2020}]{yamamoto2020parallel}
Ryuichi Yamamoto, Eunwoo Song, and Jae-Min Kim.
\newblock Parallel {WaveGAN}: A fast waveform generation model based on generative adversarial networks with multi-resolution spectrogram.
\newblock In {\em Proc. ICASSP}, pages 6199--6203. IEEE, 2020.

\bibitem[\protect\citeauthoryear{Yu \bgroup \em et al.\egroup }{2023}]{yu23b_interspeech}
Jianwei Yu, Yi~Luo, Hangting Chen, Rongzhi Gu, and Chao Weng.
\newblock High {F}idelity {S}peech {E}nhancement with {B}and-split {RNN}.
\newblock In {\em Proc. Interspeech}, pages 2483--2487, 2023.

\bibitem[\protect\citeauthoryear{Zen \bgroup \em et al.\egroup }{2019}]{zen19_interspeech}
Heiga Zen, Viet Dang, Rob Clark, Yu~Zhang, Ron~J Weiss, Ye~Jia, Zhifeng Chen, and Yonghui Wu.
\newblock {LibriTTS}: A {C}orpus {D}erived from {LibriSpeech} for {Text-to-Speech}.
\newblock In {\em Proc. Interspeech}, pages 1526--1530, 2019.

\bibitem[\protect\citeauthoryear{Zhou \bgroup \em et al.\egroup }{2024}]{zhou2024mel}
Rui Zhou, Xian Li, Ying Fang, and Xiaofei Li.
\newblock Mel-{F}ull{S}ub{N}et: Mel-{S}pectrogram {E}nhancement for {I}mproving {B}oth {S}peech {Q}uality and {ASR}.
\newblock {\em arXiv preprint arXiv:2402.13511}, 2024.

\end{thebibliography}

\end{document}